\documentclass[journal]{IEEEtran}

\usepackage{amsfonts}
\usepackage{amsmath}
\usepackage{amssymb}
\usepackage{graphicx}
\usepackage[nonscript]{accents}%
\providecommand{\U}[1]{\protect\rule{.1in}{.1in}}

\begin{document}

\title{Squeezing via spontaneous rotational symmetry breaking in a four--wave mixing
cavity}
%
%
%

\author{Ferran V. Garcia--Ferrer, Carlos Navarrete--Benlloch, Germ\'{a}n J. de
Valc\'{a}rcel and Eugenio Rold\'{a}n
\thanks{All the authors are with Departament d'\`{O}ptica, Universitat de Val\`{e}ncia, Dr. Moliner 50,
46100--Burjassot, Spain. (e-mail: carlos.navarrete@uv.es)}
\thanks{The authors want to thank Alberto Aparici for his help with the final \LaTeX~code.}
\thanks{This work has been supported by the Spanish Ministerio de Ciencia e
Innovaci\'{o}n and the European Union FEDER through Projects FIS2005-07931-C03-01 and
FIS2008-06024-C03-01. C. Navarrete-Benlloch is grant holder of the
Programa FPU del Ministerio de Ciencia e Innovaci\'{o}n.}
\thanks{This work has been submitted to the IEEE for possible
publication. Copyright may be transferred without notice, after
which this version may no longer be accessible.}%
}

%
%

\markboth{}%
{}
%



\maketitle

\begin{abstract}
We predict the generation of noncritically squeezed light through the
spontaneous rotational symmetry breaking occurring in a Kerr cavity. The model
considers a $\chi^{(3)}$ cavity that is pumped by two Gaussian beams of
frequencies $\omega_{1}$ and $\omega_{2}$. The cavity configuration is such
that two signal modes of equal frequency $\omega_{\mathrm{s}}=\left(  \omega_{1}%
+\omega_{2}\right)  /2$ are generated, these signal fields being first order
Laguerre--Gauss modes. In this system a spontaneous breaking of the rotational
symmetry occurs as the signal field corresponds to a Hermite--Gauss TEM mode.
This symmetry breaking leads to the perfect and non--critical (i.e., non
dependent on the parameter values) squeezing of the angular momentum of the
output TEM mode, which is another TEM mode spatially orthogonal to that in
which bright emission occurs.
\end{abstract}



%

\section{Introduction}

Squeezed light is a kind of radiation exhibiting reduced fluctuations with
respect to vacuum in some special observable. This occurs at the obvious
expense of an increase in the fluctuations of its canonical pair, as followed
by the Heisenberg uncertainty relation satisfied by the couple. In a single
mode field these canonically related observables correspond to orthogonal
field quadratures, which are equivalent to the position and momentum of a
harmonic oscillator. Squeezing is a macroscopic manifestation of quantum
phenomena that is attracting continuous attention since the late seventies of
the past century \cite{Loudon87,WM,Drummond04}. Nowadays a renewed interest
has arised because of the importance of squeezing in generating continuous
variable entanglement, which is a central issue for continuous variable
quantum information purposes \cite{Braunstein05}.

Squeezed light is generated by means of nonlinear optical processes, such as
parametric down--conversion or four--wave mixing. The squeezing level
attainable in such nonlinear optical processes depends on the interaction time
that is limited by the nonlinear medium length. Thus in order to increase the
squeezing level, these processes are usually confined to occur within an
optical cavity. In this way the squeezing level can reach the largest possible
levels at the system bifurcation points such as, e.g., at the emission
threshold. Squeezing levels as large as 90\% (10dB reduction respect to vacuum
fluctuations) have been recently reported \cite{10dB, Takeno07} in such conditions in
degenerate optical parametric oscillators (DOPOs). However perfect squeezing
(i.e., the complete suppression of quantum fluctuations in a field
observable)\ cannot be achieved in these conditions because complete
suppression of fluctuations in a mode quadrature implies the existence of
infinite fluctuations in the other quadrature, what would require infinite
energy in the process.

Nevertheless perfect squeezing could be actually produced and we have recently
proposed a way for obtaining it \cite{PRL}. The idea can be put in short as
follows. Consider a nonlinear optical process in which two photons with equal
frequency are generated, each photon corresponding to $\pm1$ orbital angular
momentum (OAM) Laguerre--Gauss mode. This is equivalent to generating two
photons in a TEM$_{10}$ Hermite--Gauss mode whose orientation in the
transverse x-y plane is determined by the phase difference between the two
Laguerre--Gauss photons, let us denote it by $\phi$. Now assume that $\phi$ is
not fixed as it occurs, e.g., in a down--conversion process. This amounts to
saying that the orientation of the Hermite--Gauss mode is not fixed as $\phi$
is the angle formed by the Hermite--Gauss mode with respect to the x-axis. In
these conditions we can expect the occurrence of arbitrarily large
fluctuations in the Hermite--Gauss mode orientation, which suggests that the
canonical pair of $\phi$, namely the angular momentum $-i\partial/\partial
\phi$, could be perfectly fixed. But the angular momentum of a TEM$_{10}$ mode
forming an angle $\phi$ with respect to the x-axis is another TEM$_{10}$ mode
forming an angle $\phi+\pi/2$ with respect to the x-axis. Then this mode could
exhibit perfect squeezing in one of its field quadratures. Notice that the
concept of bifurcation is not involved in this discussion and that the
variable exhibiting arbitrary fluctuations is an angle. Then, a priori,
perfect squeezing is possible in such a process as "infinite" fluctuations are
possible in the fluctuations of $\phi$.

In Refs. \cite{PRL, Largo} we have recently theoretically demonstrated the
above ideas in a model of DOPO tuned to the first transverse family at the
down-converted frequency. The requirement that the angle $\phi$ can take any
possible value (i.e. that the phase difference between the two Laguerre--Gauss
modes be arbitrary) is nothing but the requirement that the system be
rotationally invariant around the optical cavity axis. Hence the resulting
squeezing can be understood too as the result of the spontaneous breaking of
this rotational symmetry, as the emitted Hermite--Gauss mode is obviously no
more rotationally invariant. In \cite{Family} we extended this study to DOPOs
having different transverse families resonating at the down-converted
frequency, arriving to the same conclusion: Every time the nonlinear process
generates light which breaks the rotational invariance of the system, the
expected perfectly squeezed observable is found.%

\begin{figure}[t]

\includegraphics[
height=2.1914in,
width=3.0372in
]%
{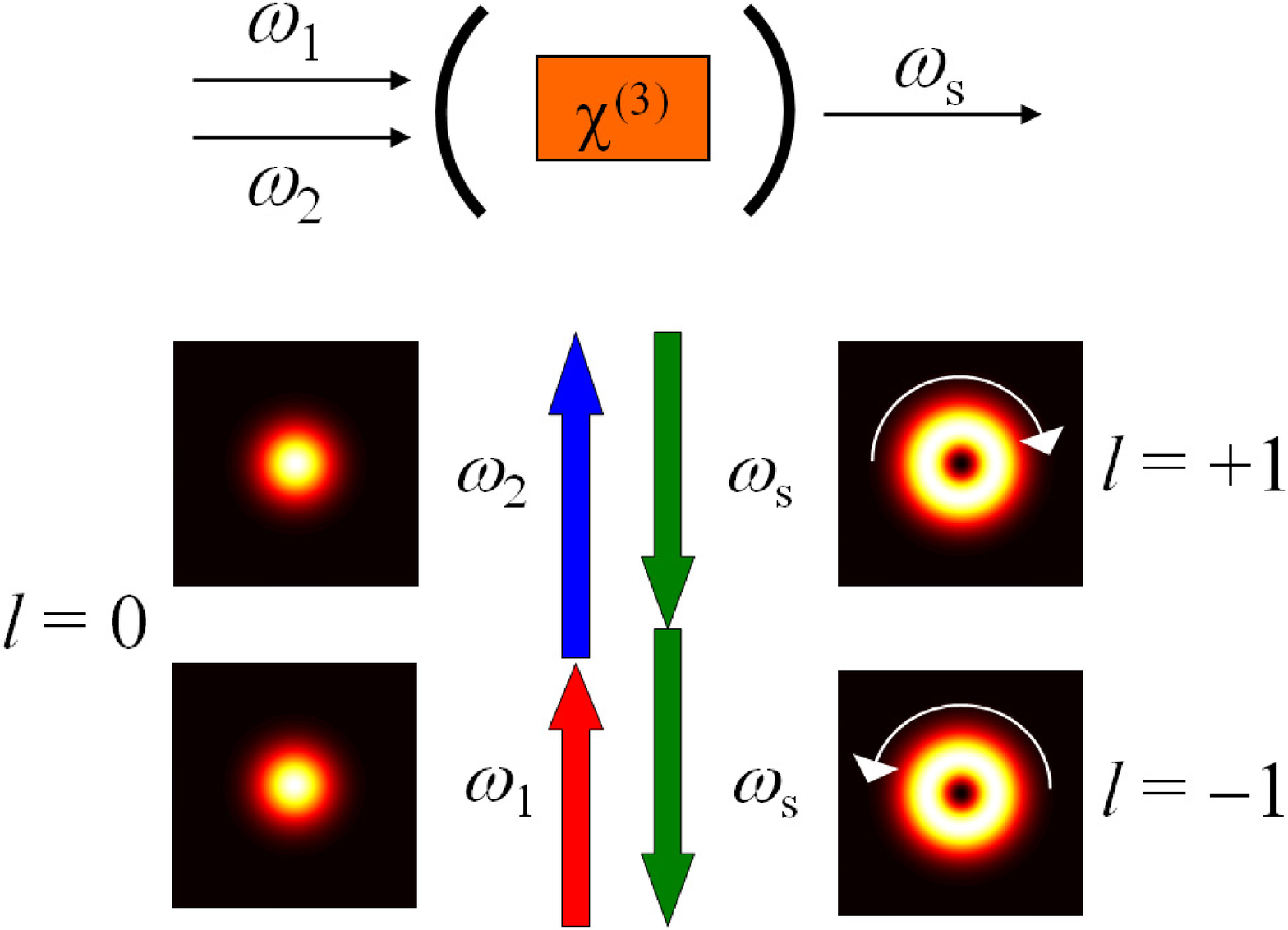}%
\\
{\protect\small Figure 1.- Scheme of the system. A $\chi^{\left(  3\right)  }$ medium is
confined within an optical cavity and pumped by two Gaussian beams of
frequencies $\omega_{1}$ and $\omega_{2}$. The cavity tuning is such that two
signal modes with frequency $\omega_{\mathrm{s}}=\left(  \omega_{1}+\omega_{2}\right)
/2$ are generated. The two signal modes are degenerated in frequency but
differ in the spatial mode, one (the other) corresponding to a Laguerre--Gauss
mode with orbital angular momentum $l=+1$ ($l=-1$).}%

\end{figure}

In the present paper we present a model for a $\chi^{\left(  3\right)  }%
$--nonlinear cavity in which squeezing appears as the result of the rotational
symmetry breaking. The interest of this new proposal is twofold. On one hand
it allows us to demonstrate that rotational symmetry breaking is a robust
means for generating squeezing in the sense that is not limited to a
$\chi^{\left(  2\right)  }$--nonlinear cavity such as the DOPO. On the other
hand perfect rotational invariance could be problematic to achieve in
$\chi^{\left(  2\right)  }$ systems because phase--matching requirements could
imply the tilting of the nonlinear crystal thus compromising rotational
invariance, a difficulty that disappears in a $\chi^{\left(  3\right)  }$
process because phase--matching occurs easily in this case.

The type of $\chi^{\left(  3\right)  }$--nonlinear cavity system we are
proposing here is a novel one that has not been studied previously, as far as
we know. Hence we must derive the quantum model (Section II) as well as study
its classical emission properties (Section III) before addressing its quantum
properties (Section IV). We are able to demonstrate that the proposed device
effectively exhibits perfect squeezing originating in the rotational symmetry
breaking. In Section V we resume our main results.

\section{Model}

Consider an optical cavity with spherical mirrors containing an isotropic
$\chi^{\left(  3\right)  }$ medium. The cavity is pumped from the outside with
two coherent fields of frequencies $\omega_{1}$ and $\omega_{2}$, these
pumping beams having a Gaussian transverse profile. Suppose, for simplicity,
that these pumping beams have the frequencies and shapes corresponding to two
consecutive longitudinal modes of the optical cavity. Then, within the cavity
the nonlinear interaction generates, through a four--wave mixing (FWM)
process, two other fields having the same frequency $\omega_{\mathrm{s}}$ such
that $\omega_{1}+\omega_{2}=2\omega_{\mathrm{s}}$. Assume now that the cavity
geometry and tuning is such that these two signal fields have the shape of
first order Laguerre--Gauss modes. These modes carry OAM and its conservation
imposes that one of the signal fields carries positive OAM with $l=+1$ while
the other carries negative OAM with $l=-1$ as the pumping fields have zero OAM
(see Fig. 1).

As stated, the just described FWM process requires that the optical cavity
modes, as well as the fields' frequencies, be properly chosen. An immediate
choice that verifies the previous requirements is a confocal resonator. In
this type of cavity the resonance frequency of longitudinal mode $q$
corresponding to the transverse family $f=2p+l$ ($p$ is the radial index) is
given by \cite{Hodgson05}%
\begin{equation}
\omega_{qf}=\frac{\pi c}{L}\left(  q+\frac{1+f}{2}\right)  ,
\end{equation}
where $L$ is the effective cavity length. In this case, the pumping beams can
correspond to two consecutive longitudinal modes with $f=0$. The signal modes
would then correspond to the cavity modes with indices $q$ and $f=1$, as they
verify $2\omega_{\mathrm{s}}=\omega_{q,0}+\omega_{q+1,0}=2\omega_{q,1}$.
Certainly, in the confocal resonator there are other modes with frequency
$\omega_{\mathrm{s}}$ (an infinite number indeed) having larger odd angular
momenta and belonging to other families. However we can neglect them by
considering that these higher order Laguerre--Gauss modes could have larger
cavity losses (what is true for low Fresnel number cavities) and would
consequently not be amplified. Once we have shown that the FWM process we
propose could be experimentally implemented, we pass to formulate the
mathematical model of our system.

\subsection{The fields}

We shall assume for simplicity that the $\chi^{\left(  3\right)  }$ crystal is
placed at the cavity's waist plane and that is thin enough as to perform the
uniform field approximation, hence neglecting any dependence of the fields on
the axial coordinate $z$. Thus we write the total quantum field inside the
cavity, at the beam waist,\ as
\begin{subequations}
\label{quantum-field}%
\begin{align}
\hat{E}\left(  \mathbf{r},t\right)   &  =\hat{E}_{\mathrm{p}}\left(
\mathbf{r},t\right)  +\hat{E}_{\mathrm{s}}\left(  \mathbf{r},t\right)  ,\\
\hat{E}_{\mathrm{p}}\left(  \mathbf{r},t\right)   &  =\underset{j=1,2}{%
{\displaystyle\sum}
}i\mathcal{F}_{j}\hat{A}_{j}\left(  \mathbf{r},t\right)  e^{-i\omega_{j}%
t}+H.c.,\\
\hat{E}_{\mathrm{s}}\left(  \mathbf{r},t\right)   &  =i\mathcal{F}%
_{\mathrm{s}}\hat{A}_{\mathrm{s}}\left(  \mathbf{r},t\right)  e^{-i\omega
_{s}t}+H.c.,
\end{align}
where $H.c.$ stands for Hermitian conjugate; $\mathbf{r=}r\left(  \cos
\phi,\sin\phi\right)  $ is the position vector in the transverse plane written
in polar coordinates; subindices $\mathrm{p}$ and $\mathrm{s}$ denote pump and
signal modes, respectively; $\mathcal{F}_{k}^{2}=\hbar\omega_{k}/\left(
\varepsilon_{0}nL\right)  $, with $k=1,2,\mathrm{s}$; and $n$ is the
refractive index (we neglect dispersion for simplicity). The slowly varying
amplitudes are
\end{subequations}
\begin{subequations}
\label{envolventL}%
\begin{align}
\hat{A}_{j}\left(  \mathbf{r},t\right)   &  =\hat{a}_{j}\left(  t\right)
G_{j}\left(  \mathbf{r}\right)  ,\ \ j=1,2,\\
\hat{A}_{\mathrm{s}}\left(  \mathbf{r},t\right)   &  =\hat{a}_{+}\left(
t\right)  L_{+}\left(  \mathbf{r}\right)  +\hat{a}_{-}\left(  t\right)
L_{-}\left(  \mathbf{r}\right)  ,
\end{align}
with $\hat{a}_{k}\left(  t\right)  $ and $\hat{a}_{k}^{\dag}\left(  t\right)
$ the annihilation and creation operators for mode $k=1,2,+,-$, which verify
$\left[  \hat{a}_{m}\left(  t\right)  ,\hat{a}_{n}^{\dagger}\left(  t\right)
\right]  =\delta_{mn}$. As for the spatial dependence in (\ref{envolventL}),
they are given by \cite{Hodgson05}%
\end{subequations}
\begin{subequations}
\begin{align}
G_{j}\left(  \mathbf{r}\right)   &  =\sqrt{\frac{2}{\pi}}\frac{1}{w_{j}%
}e^{-\left(  r/w_{j}\right)  ^{2}},\text{ }j=1,2,\\
L_{\pm1}\left(  \mathbf{r}\right)   &  =\frac{2}{\sqrt{\pi}}\frac
{r}{w_{\mathrm{s}}^{2}}e^{-\left(  r/w_{\mathrm{s}}\right)  ^{2}}e^{\pm i\phi
},
\end{align}
for the Gaussian and first order Laguerre--Gauss modes, respectively. In all
cases $w_{j}\propto1/\sqrt{\omega_{j}}$, $j=1,2,\mathrm{s}$. Notice however
that in the optical domain, in which $\omega_{1}\sim10^{15}\mathrm{%
\operatorname{s}%
}^{-1}$, one can safely take $\omega_{1}\simeq\omega_{2}\simeq\omega
_{\mathrm{s}}$ as far as $L$ is not very small, which implies that the waist
is very nearly the same for all of the involved modes. We make this
approximation that, although not essential, simplifies some expressions below.

For later use we need the relation between the Laguerre--Gauss modes and the
Hermite--Gauss modes
\end{subequations}
\begin{subequations}
\label{h's}%
\begin{align}
H_{c}^{\sigma} &  =\frac{e^{-i\sigma}L_{+1}+e^{i\sigma}L_{-1}}{\sqrt{2}}%
=\sqrt{2}\left\vert L_{\pm1}\left(  \mathbf{r}\right)  \right\vert \cos\left(
\phi-\sigma\right)  ,\label{hc}\\
H_{s}^{\sigma} &  =\frac{e^{-i\sigma}L_{+1}-e^{i\sigma}L_{-1}}{i\sqrt{2}%
}=\sqrt{2}\left\vert L_{\pm1}\left(  \mathbf{r}\right)  \right\vert
\sin\left(  \phi-\sigma\right)  ,\label{hs}%
\end{align}
being $H_{c}^{\sigma}$ and $H_{s}^{\sigma}$ the Hermite--Gauss modes with an
orientation $\sigma$ and $\sigma+\pi/2$ with respect to the \textrm{x}--axis,
respectively. Thus the slowly varying amplitudes at frequency $\omega_{s}$ can
also be written as%
\end{subequations}
\begin{equation}
\hat{A}_{\mathrm{s}}\left(  \mathbf{r},t\right)  =\hat{a}_{c,\sigma}\left(
t\right)  H_{c}^{\sigma}+\hat{a}_{s,\sigma}\left(  t\right)  H_{s}^{\sigma},
\end{equation}
with
\begin{subequations}
\label{ac,as}%
\begin{align}
\hat{a}_{c,\sigma} &  =\frac{1}{\sqrt{2}}\left(  e^{i\sigma}\hat{a}%
_{+}+e^{-i\sigma}\hat{a}_{-}\right)  ,\\
\hat{a}_{s,\sigma} &  =\frac{i}{\sqrt{2}}\left(  e^{i\sigma}\hat{a}%
_{+}-e^{-i\sigma}\hat{a}_{-}\right)  ,
\end{align}
the annihilation operators for the Hermite-Gauss modes. Finally, we introduce
the field quadratures of these Hermite--Gauss modes
\end{subequations}
\begin{equation}
\hat{X}_{j,\sigma}^{\varphi}=e^{-i\varphi}\hat{a}_{j,\sigma}+e^{i\varphi}%
\hat{a}_{j,\sigma}^{\dagger},\ \ j=c,s,\label{q's}%
\end{equation}
with $\hat{a}_{c,\sigma}$ and $\hat{a}_{s,\sigma}$ given by Eqs. (\ref{ac,as}).

\subsection{The Hamiltonian}

In the interaction picture, the system's Hamiltonian can be written as%
\begin{equation}
\hat{H}=\hat{H}_{0}+\hat{H}_{\mathrm{ext}}+\hat{H}_{\mathrm{int}}.
\end{equation}
$\hat{H}_{0}$ and $\hat{H}_{ext}$ correspond to the modes' energies and
external injection, respectively, and are given by
\begin{subequations}
\begin{align}
\hat{H}_{0}  &  =\sum_{j=1,2,+,-}\hbar\delta_{j}a_{j}^{\dagger}a_{j},\\
\hat{H}_{\mathrm{ext}}  &  =i\hbar\mathcal{E}_{1}\left(  \hat{a}_{1}^{\dagger
}-\hat{a}_{1}\right)  +i\hbar\mathcal{E}_{2}\left(  \hat{a}_{2}^{\dagger}%
-\hat{a}_{2}\right)  ,
\end{align}
with $\delta_{j}=\left(  \omega_{Cj}-\omega_{j}\right)  $ the cavity detuning
for the mode with frequency $\omega_{j}$, being $\omega_{Cj}$ the cavity
resonance closest to that mode. In a confocal resonator this detuning is the
same for all the modes if the relative frequency of the pump modes is locked
to the free spectral range of the cavity, i.e., $\omega_{2}-\omega_{1}=\pi
c/L$. Hence, in the following we take $\delta_{j}=\delta$ $\forall$ $j$.
$\mathcal{E}_{j}$ are the pumping parameters, which are related to the
experimental parameters by
\end{subequations}
\[
\mathcal{E}_{j}=\sqrt{\frac{c}{2\omega_{j}\hbar L}T\left(  \omega_{j}\right)
P_{j}},
\]
being $T\left(  \omega_{j}\right)  $ the transmission factor at the considered
frequency and $P_{j}$ the power of the pumped laser. In the following we will
assume $\mathcal{E}_{1}=\mathcal{E}_{2}=\mathcal{E}$ for simplicity. $\hat
{H}_{int}$ describes the nonlinear interaction and can be written as the sum
of three contributions%

\begin{equation}
\hat{H}_{\mathrm{int}}=-\hbar g\left(  \hat{H}_{\mathrm{spm}}+\hat
{H}_{\mathrm{cpm}}+\hat{H}_{\mathrm{fwm}}\right)  ,
\end{equation}
with%
\begin{align}
\hat{H}_{\mathrm{spm}}  &  =\hat{a}_{1}^{\dagger2}\hat{a}_{1}^{2}+\hat{a}%
_{2}^{\dagger2}\hat{a}_{2}^{2}+\frac{1}{2}\left(  \hat{a}_{+}^{\dagger2}%
\hat{a}_{+}^{2}+\hat{a}_{-}^{\dagger2}\hat{a}_{-}^{2}\right)  ,\\
\hat{H}_{\mathrm{cpm}}  &  =4\hat{a}_{1}^{\dagger}\hat{a}_{1}\hat{a}%
_{2}^{\dagger}\hat{a}_{2}+2\hat{a}_{+}^{\dagger}\hat{a}_{+}\hat{a}%
_{-}^{\dagger}\hat{a}_{-}\\
&  +2\left(  \hat{a}_{1}^{\dagger}\hat{a}_{1}+\hat{a}_{2}^{\dagger}\hat{a}%
_{2}\right)  \left(  \hat{a}_{+}^{\dagger}\hat{a}_{+}+\hat{a}_{-}^{\dagger
}\hat{a}_{-}\right)  ,\nonumber\\
\hat{H}_{\mathrm{fwm}}  &  =2\left(  \hat{a}_{1}^{\dagger}\hat{a}_{2}%
^{\dagger}\hat{a}_{+}\hat{a}_{-}+\hat{a}_{1}\hat{a}_{2}\hat{a}_{+}^{\dagger
}\hat{a}_{-}^{\dagger}\right)  ,
\end{align}
describing self--phase modulation, cross--phase modulation, and four--wave
mixing, respectively. Note that this Hamiltonian contains all the possible
combinations of four operators conserving both energy and OAM. The factors
multiplying the different terms are intuitive once one takes into account the
two following features: (i),\ there is a global factor 4 in $\hat
{H}_{\mathrm{cpm}}$ and $\hat{H}_{\mathrm{fwm}}$ with respect to $\hat
{H}_{\mathrm{spm}}$ coming from all possible permutations of the different
operators; and (ii), if any of the four modes is a signal mode, a factor 1/2
appears that comes from the transverse modes' overlapping integral. Finally,
it can be shown that the coupling constant is given by%
\begin{equation}
g=\frac{6\mathcal{F}^{4}\varepsilon_{0}l_{\mathrm{c}}\chi}{\pi\hbar w^{2}},
\label{g}%
\end{equation}
with $w$ the beam waist, $l_{\mathrm{c}}$ the crystal length and $\chi$ the
third-order nonlinear susceptibility of the crystal.

\subsection{The quantum evolution equations}

In this subsection we apply the standard procedure to develop the quantum
theory of a nonlinear resonator within the generalized $P$ representation to
our Kerr cavity model \cite{Drummond80,Carmichael,Gardiner00}. The starting
point is the system's master equation for the density operator $\hat{\varrho}%
$, which reads%
\begin{equation}
\frac{d}{dt}\hat{\varrho}=\frac{1}{i\hbar}\left[  \hat{H},\hat{\varrho
}\right]  +\widehat{\mathcal{L}\varrho}, \label{master}%
\end{equation}
where $\mathcal{L}$ is the Liouvillian superoperator describing field losses
through the output mirror, which applied to the density operator reads%
\begin{equation}
\widehat{\mathcal{L}\varrho}=\sum_{j=1,2,+,-}\gamma_{j}\left(  \left[  \hat
{a}_{j},\hat{\varrho}\hat{a}_{j}^{\dagger}\right]  +\left[  \hat{a}_{j}%
\hat{\varrho},\hat{a}_{j}^{\dagger}\right]  \right)  .
\end{equation}
As we are assuming that the system has perfect rotational symmetry around the
cavity axis it follows that $\gamma_{+}=\gamma_{-}\equiv\gamma_{\mathrm{s}}$.

As usual, we use now the generalized $P$ representation in order to transform
the operator master equation into a partial differential equation for the
quasiprobability distribution $P$. In this representation, to every pair of
boson operators $\left(  \hat{a}_{j},\hat{a}_{j}^{\dagger}\right)  $ it
corresponds a pair of independent stochastic amplitudes $\left(  \alpha
_{j},\alpha_{j}^{+}\right)  $ verifying $\left\langle \alpha_{j}%
^{+}\right\rangle =\left\langle \alpha_{j}\right\rangle ^{\ast}$. By using
standard techniques, one finds that the Fokker--Planck equation governing the
evolution of $P$ reads%
\begin{equation}
\frac{\partial}{\partial t}P\left(  \boldsymbol{\alpha};t\right)  =\left[
-\frac{\partial}{\partial\alpha_{i}}A_{i}+\frac{1}{2}\frac{\partial^{2}%
}{\partial\alpha_{i}\partial\alpha_{j}}\mathcal{D}_{ij}\right]  P\left(
\boldsymbol{\alpha};t\right)  , \label{FP}%
\end{equation}
where we write vector $\boldsymbol{\alpha}$ as%
\begin{equation}
\boldsymbol{\alpha}=\left(  \alpha_{1},\alpha_{2},\alpha_{+},\alpha_{-}%
,\alpha_{1}^{+},\alpha_{2}^{+},\alpha_{+}^{+},\alpha_{-}^{+}\right)  ,
\end{equation}
and give the explicit expression of the components of both the drift vector
$\mathbf{A}\left(  \boldsymbol{\alpha}\right)  $ and the diffusion matrix
$\mathcal{D}\left(  \boldsymbol{\alpha}\right)  $ in Appendix A.

Once the Fokker--Planck equation is known, it can be converted, by applying
Ito rules, into an equivalent set of stochastic first-order differential
equations: The quantum Langevin equations. They read%
\begin{equation}
\frac{d}{dt}\boldsymbol{\alpha}=\mathbf{A}\left(  \boldsymbol{\alpha}\right)
+\mathcal{B}\left(  \boldsymbol{\alpha}\right)  \cdot\boldsymbol{\eta}\left(
t\right)  \label{lang}%
\end{equation}
where the components of $\boldsymbol{\eta}\left(  t\right)  $ are real
gaussian noises verifying
\begin{subequations}
\label{noise}%
\begin{align}
\left\langle \eta_{i}\left(  t\right)  \right\rangle  &  =0,\\
\left\langle \eta_{i}\left(  t\right)  \eta_{j}\left(  t^{\prime}\right)
\right\rangle  &  =\delta_{ij}\delta\left(  t-t^{\prime}\right)  ,
\end{align}
and the noise matrix $\mathcal{B}$ can be obtained from%
\end{subequations}
\begin{equation}
\mathcal{D}=\mathcal{B}\cdot\mathcal{B}^{T}.
\end{equation}

The equivalency between the master equation and Langevin equations must be
understood as follows%
\begin{equation}
\left\langle :f\left(  \hat{a}_{m},\hat{a}_{m}^{\dagger}\right)
:\right\rangle =\left\langle f\left(  \alpha_{m},\alpha_{m}^{+}\right)
\right\rangle _{stochastic}, \label{N-St}%
\end{equation}
i.e., quantum expected values of normally ordered functions are equal to the
stochastic averages of the same functions after changing boson operators
$\left(  \hat{a}_{m},\hat{a}_{m}^{\dagger}\right)  $ by independent complex
stochastic variables $\left(  \alpha_{m},\alpha_{m}^{+}\right)  $.

With this procedure, we have obtained a set of stochastic differential
equations, Eqs. (\ref{lang}) together with Eqs. (\ref{Alan}) and (\ref{Dlan}),
ruling the evolution of the fields inside the cavity. However, the analysis of
this model turns out to be quite involved.

A simpler model retaining the essential ingredients consists in neglecting the
temporal evolution of the pumping fields, or further, considering the
undepletion limit for the pump. As follows from (\ref{A1}) and (\ref{A2})
applied to (\ref{lang}), this will be a reasonable approximation whenever the
gain $g$, Eq. (\ref{g}), is small enough as for $\left\vert \alpha_{\pm
}\right\vert \ll$ $\left\vert \alpha_{1,2}\right\vert $. In such a case one
can safely neglect the amplitude variation of the pump modes. Of course, in
doing that some of the dynamical richness of the system is lost, but the
approximation simplifies very much the analysis of the rotational symmetry
breaking which is fully retained by the simplified model.

Hence, in the following we will study a reduced model in which the pump fields
are taken to be equal (and real without loss of generalization), i.e., we take%
\begin{equation}
\alpha_{1}=\alpha_{2}\equiv\rho\in%
\mathbb{R}
. \label{RedAssump}%
\end{equation}

\subsection{The reduced model}

Under the assumption that the pump fields remain constant, thus satisfying
(\ref{RedAssump}), the Fokker-Planck equation of the system looks like Eq.
(\ref{FP}), but with simpler diffusion matrix and drift vector.

From the general expressions given in Appendix A it is easy to obtain that the
diffusion matrix reads now
\begin{subequations}
\label{D}%
\begin{align}
\mathcal{D}  &  =%
\begin{pmatrix}
\mathcal{D}^{\left(  -\right)  } & 0\\
0 & \mathcal{D}^{\left(  +\right)  }%
\end{pmatrix}
,\label{Dtotal}\\
\mathcal{D}^{\left(  -\right)  }  &  =ig%
\begin{pmatrix}
\alpha_{+}^{2} & 2\alpha_{+}\alpha_{-}+2\rho^{2}\\
2\alpha_{+}\alpha_{-}+2\rho^{2} & \alpha_{-}^{2}%
\end{pmatrix}
,
\end{align}
($\mathcal{D}^{\left(  +\right)  }$ is like $\mathcal{D}^{\left(  -\right)  }$
but swapping $\alpha_{k}\ $and $\alpha_{k}^{+}$, and changing $i$ by $-i$),
and that the components of the drift vector are
\end{subequations}
\begin{align}
A_{\alpha_{\pm}}=-\left(  \gamma_{s}+i\delta\right)  \alpha_{\pm}  &
+i\left(  \alpha_{\pm}^{+}\alpha_{\pm}+2\alpha_{\mp}^{+}\alpha_{\mp}+4\rho
^{2}\right)  \alpha_{\pm}\nonumber\\
&  +2i\rho^{2}\alpha_{\mp}^{+},
\end{align}
being $A_{\alpha_{\pm}^{+}}$ like $A_{\alpha_{\pm}}$ after swapping
$\alpha_{k}\ $and $\alpha_{k}^{+}$, and changing $i$ by $-i$.

Like before, from this Fokker-Planck equation, a set of Langevin equations can
be obtained. Nevertheless, it will be useful to introduce the following change
of variables and parameters (we note that the phase factor $\psi$ appearing in
the rescaled fields is introduced to make the stationary
solutions of the classical equations satisfy $\bar{\beta}_{-}=\bar{\beta}%
_{+}^{\ast}$, see Eqs. (\ref{solON}), which simplifies the quantum analysis)
\begin{align}
\ \beta_{\pm}  &  =\sqrt{\frac{g}{\gamma_{s}}}\alpha_{\pm}e^{-i\psi},\text{
}\beta_{\pm}^{+}=\sqrt{\frac{g}{\gamma_{s}}}\alpha_{\pm}^{+}e^{i\psi
},\label{reescal}\\
\text{ }p  &  =2\frac{g}{\gamma_{s}}\rho^{2},\ \ \Delta=\frac{\delta}%
{\gamma_{s}},\ \ \kappa=\frac{g}{\gamma_{s}},\text{ }T=\gamma_{s}t\nonumber
\end{align}
with
\begin{equation}
\sin2\psi=\sqrt{\frac{\gamma_{s}}{g}}\frac{1}{2\rho}=\frac{1}{\sqrt{2p}},
\label{psi}%
\end{equation}
in terms of which the Langevin equations for the reduced model read%
\begin{align}
\dot{\beta}_{\pm}  &  =-\left(  1+i\Delta\right)  \beta_{\pm}+i\left(
\beta_{\pm}^{+}\beta_{\pm}+2\beta_{\mp}^{+}\beta_{\mp}+2p\right)  \beta_{\pm
}\nonumber\\
&  +ipe^{-2i\psi}\beta_{\mp}^{+}+\mathcal{\tilde{B}}_{\beta_{\pm},j}\xi
_{j}\left(  T\right)  , \label{L}%
\end{align}
plus the corresponding equations for $\beta_{\pm}^{+}$ which are like those
for $\beta_{\pm}$ above after swapping $\beta_{k}\ $and $\beta_{k}^{+}$, and
changing $i$ by $-i$. The overdot indicates derivative with respect to the
adimensional time $T$, and the four components of the noise vector
$\boldsymbol{\xi}$ satisfy properties (\ref{noise}) now for the adimensional
time $T$. As for the noise matrix $\mathcal{\tilde{B}}$, it can be obtained
from $\mathcal{\tilde{D}}=\mathcal{\tilde{B}}\cdot\mathcal{\tilde{B}}^{T}$ as
usual, but now using the diffusion matrix after introducing the changes
(\ref{reescal}), which reads as (\ref{Dtotal}) but with%
\[
\mathcal{\tilde{D}}^{\left(  -\right)  }=ig%
\begin{pmatrix}
\alpha_{+}^{2} & 2\alpha_{+}\alpha_{-}+2e^{-2i\psi}\rho^{2}\\
2\alpha_{+}\alpha_{-}+2e^{-2i\psi}\rho^{2} & \alpha_{-}^{2}%
\end{pmatrix}
,
\]
and with $\mathcal{D}^{\left(  +\right)  }$ like $\mathcal{D}^{\left(
-\right)  }$ but swapping $\alpha_{k}\ $and $\alpha_{k}^{+}$, and changing $i$
by $-i$. In any case, there is no need to evaluate $\mathcal{\tilde{B}}$ at
this moment.

Eqs. (\ref{L}) constitute the model we analyze in detail below.

\section{Classical limit}

Before addressing the analysis of quantum fluctuations we need to know the
classical steady states of the system as well as their stability properties,
and for doing that we must first write down the classical limit of Eqs.
(\ref{L}). This is easily done by neglecting the noise terms and by making
$\beta_{j}^{+}=\beta_{j}^{\ast}$ in the quantum Langevin Eqs. (\ref{L}). We
obtain the following set of two complex ordinary differential equations for
the signal classical fields amplitudes $\beta_{\pm}$:%

\begin{align}
\dot{\beta}_{\pm}  &  =-\left[  1+i\left(  \Delta-\left\vert \beta_{\pm
}\right\vert ^{2}-2\left\vert \beta_{\mp}\right\vert ^{2}-2p\right)  \right]
\beta_{\pm}\nonumber\\
&  +ipe^{-2i\psi}\beta_{\mp}^{\ast}. \label{cla2}%
\end{align}
Thanks to rescaling (\ref{reescal}) we can appreciate that the classical
dynamics of the system is governed by just two parameters, namely the
normalized detuning $\Delta$ and pump strength $p$. We pass now to study the
stationary solutions of Eqs. (\ref{cla2}) as well as their stability properties.

Eqs. (\ref{cla2}) have two steady states. First there is the trivial steady
state $\beta_{\pm}=0$. It is easy to show that its stability is governed by
the eigenvalues%
\begin{equation}
\lambda_{\pm}=-1\pm\sqrt{p^{2}-\left(  \Delta-2p\right)  ^{2}},
\end{equation}
what implies that the trivial solution is stable except when the pump
amplitude $p$ verifies $p_{-}<p<p_{+}$ with
\begin{equation}
p_{\pm}=\frac{1}{3}\left(  2\Delta\pm\sqrt{\Delta^{2}-3}\right)  , \label{p+-}%
\end{equation}
in which case the trivial solution becomes linearly unstable because
$\operatorname{Re}\left(  \lambda_{+}\right)  >0$. Notice that a prerequisite
for the destabilization of the trivial solution is that $\Delta>\sqrt{3}$ ($p$
is positive).%

\begin{figure}[ht]

\includegraphics[
height=4.1926in,
width=3.1695in
]%
{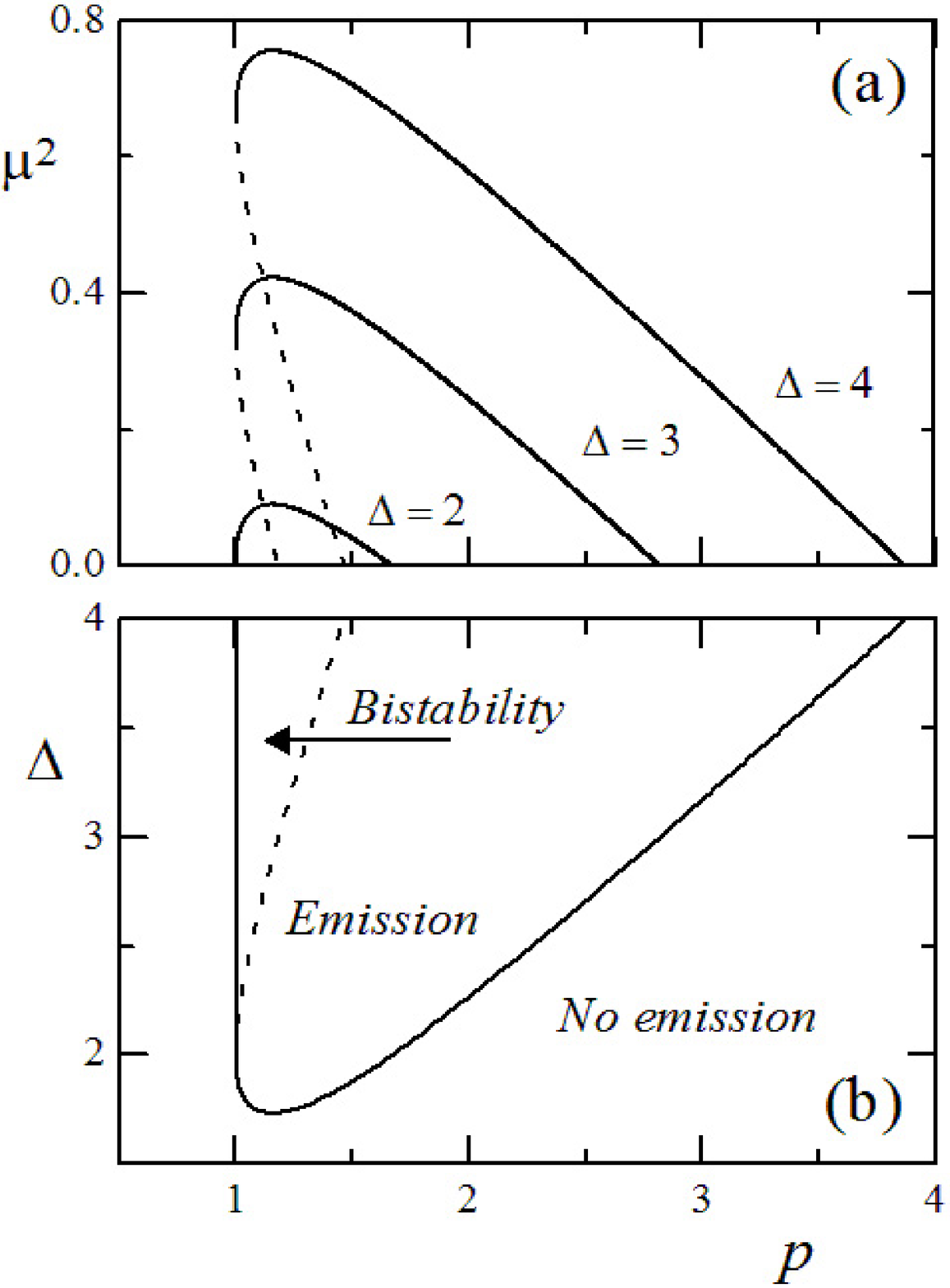}%
\\
{\protect\small Figure 2.- (a) Intensity of the emitted classical signal beam as
a function of the pump strength $p$ for
the indicated values of the detuning $\Delta$%
. Continuous (dashed) lines indicate linearly stable
(unstable) solutions. (b) Domain of existence and stability of the solutions.
The non trivial solution exists in the domain limited by the two continuous
lines. The trivial solution is stable to the left of the dashed line and to
the right of the right continuous line. Hence, the system is bistable in the
region between the dashed line and the left continuous line, marked in the
figure.
}%

\end{figure}

At the instability points $\operatorname{Re}\left(  \lambda_{+}\right)  =0$
the system passes from the trivial state to the nontrivial one, which reads%
\begin{equation}
\bar{\beta}_{\pm}=\mu e^{\mp i\theta}, \label{solON}%
\end{equation}
with%
\begin{equation}
\mu^{2}=\frac{1}{3}\left(  \Delta-2p\pm\sqrt{p^{2}-1}\right)  , \label{mu}%
\end{equation}
and where $\theta$ is half the phase--difference between the two
Laguerre--Gauss modes amplitudes, and is not fixed by Eqs. (\ref{cla2}), being
hence arbitrary. In Fig. 2(a)\ this solution is shown as a function of $p$ for
three values of $\Delta$.

Notice first that there are two possible values for $\mu$. It is easy to
demonstrate that the solution with the minus sign in front of the square root
(dashed lines in Fig. 2(a)) is always unstable, while the solution with the
plus sign (continuous lines in Fig. 2(a)) is stable within all its domain of
existence. Notice also that while the sum of the phases of the two
Laguerre--Gauss modes amplitudes is fixed to zero (this is thanks to the
change of variables (\ref{reescal})), their phase difference $\theta$ is an
arbitrary quantity, which reflects the rotational invariance of the system as
this phase difference determines the orientation in the transverse plane of
the emitted Hermite-Gauss mode, see Eq. (\ref{Aclass}) below.

From Eq. (\ref{mu}) it follows that the necessary condition for the existence
of the nontrivial solution is $p>1$, and its domain of existence is determined
by the condition
\begin{equation}
\Delta>2p-\sqrt{p^{2}-1}.
\end{equation}
Note that this inequality can be satisfied only if $\Delta>\sqrt{3}$, in
agreement with our analysis of the trivial solution. In Fig. 2(b) we represent
this domain of existence as well as the domain of stability of the trivial
solution, see Fig. 2(b) caption.

From all the above we see that the classical nontrivial state slowly--varying
amplitude can be written as%
\begin{align}
A_{s}^{class}\left(  \mathbf{r}\right)   &  =\mu\left[  e^{-i\theta}%
L_{+}\left(  \mathbf{r}\right)  +e^{+i\theta}L_{-}\left(  \mathbf{r}\right)
\right]  \nonumber\\
&  =\left[  \frac{2}{3}\left(  \Delta-2p+\sqrt{p^{2}-1}\right)  \right]
^{1/2}H_{c}^{\theta}\left(  \mathbf{r}\right)  .\label{Aclass}%
\end{align}
This solution corresponds to a Hermite--Gauss mode rotated an angle $\theta$
with respect to the \textrm{x}--axis, being the value of this angle arbitrary.
As already noticed, the arbitrariness of $\theta$ is the consequence of the
rotational symmetry of the system: As the pumping modes are rotationally
symmetric, as the cavity is, the signal TEM$_{10}$\ Hermite-Gauss mode cannot
have any preferred orientation and thus all orientations are equally likely.
This property is essential for the results we present below for quantum fluctuations.

Finally, note that the fact that the emission takes place in a TEM$_{10}$
mode, allows us to distinguish between two different modes: The $H_{c}%
^{\theta}\left(  \mathbf{r}\right)  $ mode, which is the classically generated
one, and its orthogonal mode $H_{s}^{\theta}\left(  \mathbf{r}\right)  $,
which is empty of photons at the classical level. In the following we shall
refer to these modes as the \textit{bright} and \textit{dark} modes, respectively.

\section{Quantum analysis}

We return to the quantum description of the system. In the following
subsections we first derive the linearized Langevin equations and then solve
them with the method introduced in \cite{Perez, Perez2} and further used in \cite{PRL,
Largo, Family}. Next we show the most outstanding quantum properties of the
system: After proving that the bright mode is rotating randomly in the
transverse plane, we show that the dark mode has perfect noise reduction in
one of its quadratures.

\subsection{Linearization of the Langevin equations}

We linearize Eqs. (\ref{L}), a valid approximation in the large photon number
limit and thus appropriate for our purposes as we are going to analyze quantum
fluctuations around the above threshold solution (\ref{solON}). Consequently
we write
\begin{subequations}
\label{beta}%
\begin{align}
\beta_{\pm}  &  =\left[  \mu+b_{\pm}\left(  T\right)  \right]  e^{\mp
i\theta\left(  T\right)  },\\
\beta_{\pm}^{+}  &  =\left[  \mu+b_{\pm}^{+}\left(  T\right)  \right]  e^{\pm
i\theta\left(  T\right)  },
\end{align}
where the phase difference $\theta$ appears explicitly in order to keep the
$b$'s small. Then, by assuming that $b_{j}$, $\xi_{j}$, and $\dot{\theta}$ are
small quantities, we easily arrive to the following linearized Langevin equations%

\end{subequations}
\begin{equation}
\mathbf{\dot{b}}+iN_{0}\mu\dot{\theta}\mathbf{v}_{0}=\mathcal{L}%
\mathbf{b}+\mathcal{K\bar{B}}\boldsymbol{\xi} \label{Ll}%
\end{equation}
where%
\begin{subequations}
\begin{align}
\mathbf{b}  &  =\operatorname{col}\left(  b_{+},b_{-},b_{+}^{+},b_{-}%
^{+}\right)  ,\\
\mathbf{v}_{0}  &  =\frac{1}{N_{0}}\operatorname{col}\left(  -1,1,1,-1\right)
,
\end{align}
$N_{0}$ is a normalization factor (see Eq. (\ref{N0}) below), $\mathcal{L}$ is
a matrix with elements
\end{subequations}
\begin{equation}
\mathcal{L}_{ij}=\left.  \frac{\partial A_{i}}{\partial\beta_{j}}\right\vert
_{\boldsymbol{\beta}=\boldsymbol{\bar{\beta}}},
\end{equation}
and $\mathcal{K}$ is the diagonal matrix%
\begin{equation}
\mathcal{K}=\operatorname{diag}\left(  e^{i\theta},e^{-i\theta},e^{-i\theta
},e^{i\theta}\right)  . \label{F}%
\end{equation}
Finally, $\mathcal{\bar{B}}$ refers to matrix $\mathcal{\tilde{B}}$ evaluated
at the stationary state (\ref{solON}). Its expression can be derived from
$\mathcal{\bar{D}}=\mathcal{\bar{B}}\cdot\mathcal{\bar{B}}^{T}$ with
\begin{subequations}
\label{matrizD}%
\begin{align}
\mathcal{\bar{D}}  &  =\left.  \mathcal{\tilde{D}}\right\vert
_{\boldsymbol{\beta}=\boldsymbol{\bar{\beta}}}=\left(
\begin{array}
[c]{cc}%
\mathcal{\bar{D}}^{\left(  -\right)  } & 0\\
0 & \mathcal{\bar{D}}^{\left(  +\right)  }%
\end{array}
\right)  ,\\
\mathcal{\bar{D}}^{\left(  -\right)  }  &  =\left[  \mathcal{\bar{D}}^{\left(
+\right)  }\right]  ^{\ast}=i\kappa\left(
\begin{array}
[c]{cc}%
\mu^{2}e^{-2i\theta} & 2\mu^{2}+pe^{-2i\psi}\\
2\mu^{2}+pe^{-2i\psi} & \mu^{2}e^{2i\theta}%
\end{array}
\right)  .
\end{align}
After some algebra, it is easy to show that $\mathcal{\bar{B}}$ can be written
as
\end{subequations}
\begin{subequations}
\label{B}%
\begin{align}
\mathcal{\bar{B}}  &  \mathcal{=}\left(
\begin{array}
[c]{cc}%
\mathcal{\bar{B}}^{\left(  -\right)  } & 0\\
0 & \mathcal{\bar{B}}^{\left(  +\right)  }%
\end{array}
\right)  ,\\
\mathcal{\bar{B}}^{\left(  -\right)  }  &  =\left[  \mathcal{\bar{B}}^{\left(
+\right)  }\right]  ^{\ast}=%
\begin{pmatrix}
ae^{-i\theta} & be^{-i\theta}\\
ce^{i\theta} & de^{i\theta}%
\end{pmatrix}
,
\end{align}
where elements $\left(  a,b,c,d\right)  $ are independent of $\theta$, and
satisfy the following relations%
\end{subequations}
\begin{subequations}
\begin{align}
a^{2}+b^{2}  &  =c^{2}+d^{2}=i\kappa\mu^{2},\\
ac+bd  &  =i\kappa\left(  2\mu^{2}+pe^{-2i\psi}\right)  .
\end{align}

\subsection{Solving the linearized Langevin equations}

In order to solve the linearized Langevin Eqs. (\ref{Ll}) we follow a
procedure analogous to that in \cite{PRL, Largo, Family, Perez, Perez2}. The method
consists in projecting Eqs. (\ref{Ll}) into the eigensystem of the linear
operator $\mathcal{L}$. This provides a direct way for the evaluation of the
quantum fluctuations of the relevant physical quantities, as we show below.

Operator $\mathcal{L}$ has not an orthonormal but a biorthonormal eigensystem,
i.e., there is an eigensystem of operators $\mathcal{L}$ and $\mathcal{L}%
^{\dagger}$ verifying
\end{subequations}
\begin{subequations}
\begin{equation}
\mathcal{L}\mathbf{v}_{j}=\lambda_{j}\mathbf{v}_{j},\ \ \mathcal{L}^{\dagger
}\mathbf{w}_{j}=\lambda_{j}^{\ast}\mathbf{w}_{j},
\end{equation}
such that
\end{subequations}
\begin{equation}
\mathbf{w}_{m}^{\ast}\cdot\mathbf{v}_{n}=\delta_{mn}.
\end{equation}
The quantitative result of the analysis of $\mathcal{L}$ and $\mathcal{L}%
^{\dagger}$ is that their four eigenvalues read
\begin{subequations}
\begin{align}
\lambda_{0}  &  =0,\ \ \ \ \ \ \lambda_{1}=-2,\\
\lambda_{2}  &  =\lambda_{2}\left(  p,\Delta\right)  ,\ \ \lambda
_{3}=-2-\lambda_{2}.
\end{align}
We see that eigenvalue $\lambda_{0}$ is always null. This means that its
corresponding eigenvector, which is said to be a Goldstone mode (its
expression is given below), is neutrally stable, i.e., that its \ associated
variable can take any possible value. Of course this reflects the
indeterminacy of the phase difference between the two Laguerre--Gauss modes,
$\theta$, or what is the same, the indeterminacy of the orientation of the
Hermite--Gauss output mode in the transverse plane. Hence, the null eigenvalue
implies that the fluctuations introduced by quantum noise in this orientation
are not damped and, hence, that quantum noise will induce arbitrary rotations
of the Hermite--Gauss output mode in the transverse plane. Thus the breaking
of the rotational symmetry of the system introduced by the appearance of the
Hermite--Gauss mode is, in a sense, counteracted by quantum noise by making
possible any possible orientation. Together with $\lambda_{0}=0$, there is the
companion eigenvalue $\lambda_{1}=-2$. Its corresponding eigenvector (see
below)\ is consequently maximally damped irrespective of the system's
parameter values. As we show below, it is the observable associated to the
eigenvector corresponding to $\lambda_{1}$ the one that is perfectly squeezed.

On the other hand, the two other eigenvalues, $\lambda_{2}$ and $\lambda_{3}$,
are complex in general; nevertheless, they reach the values $0$ and $-2$ at the
bifurcation points (i.e., at the points in the parameter space separating the
regions where the steady state solution (\ref{solON}) exist or not).
Consequently, their associated eigenvectors are those exhibiting the usual
squeezing occurring only at the bifurcations. This squeezing is not perfect
(it seems perfect only owed to the linearized treatment) and degrades quickly
as the system parameters are brought apart from the bifurcation points. We
shall not analyze this squeezing here because there is not any relevant new
feature in it with respect to what has been described many times in other
nonlinear optical cavities. Hence, in the following, we concentrate on the
analysis of the modes associated to the first two eigenvalues that are the
ones connected with the rotational symmetry breaking.

It is not difficult to show that the eigenvectors of $\mathcal{L}$ associated
to $\lambda_{0}$ and $\lambda_{1}$ are%
\end{subequations}
\begin{subequations}
\begin{align}
\mathbf{v}_{0}  &  =\frac{1}{N_{0}}\operatorname{col}\left(  -1,1,1,-1\right)
,\\
\mathbf{v}_{1}  &  =\frac{1}{N_{0}}\operatorname{col}\left(  e^{i\phi_{0}%
},-e^{i\phi_{0}},e^{-i\phi_{0}},-e^{-i\phi_{0}}\right)  ,
\end{align}
with%
\end{subequations}
\begin{equation}
N_{0}=-4\cos\phi_{0}, \label{N0}%
\end{equation}
a normalization factor and $\phi_{0}$ a real quantity given by%
\begin{equation}
e^{2i\phi_{0}}=\frac{\mu^{2}+pe^{-i\psi}}{2\left(  \mu^{2}+p\right)  -\left(
\Delta+i\right)  }. \label{fi0}%
\end{equation}
As for the eigenvectors of $\mathcal{L}^{\dagger}$ they read%
\begin{subequations}
\begin{align}
\mathbf{w}_{0}  &  =\operatorname{col}\left(  e^{i\phi_{0}}%
,-e^{i\phi_{0}},-e^{-i\phi_{0}},e^{-i\phi_{0}}\right)  ,\\
\mathbf{w}_{1}  &  =\operatorname{col}\left(  1,-1,1,-1\right)  .
\end{align}

Once these eigenvectors of the linear operator are known, we proceed to
project quantum fluctuations onto them. We define projections%
\end{subequations}
\begin{equation}
c_{j}=\mathbf{w}_{i}^{\ast}\cdot\mathbf{b},\ \ j=0,1. \label{c's}%
\end{equation}
Note that these projections can be easily related with the fluctuations of the
quadratures associated to modes $H_{c}^{\sigma}$ and $H_{s}^{\sigma}$, see
Eqs. (\ref{q's}). In particular, by using (\ref{q's}) and (\ref{beta}) it is
easy to arrive at%
\begin{equation}
X_{s,\theta}^{\phi_{0}}=\frac{i}{\sqrt{2}}c_{0},\ \ \ X_{s,\theta}^{\pi
/2}=\frac{1}{\sqrt{2}}c_{1}. \label{XtoC}%
\end{equation}

Next we project the linearized Langevin Eqs. (\ref{Ll}). By multiplying them
by $\mathbf{w}_{j}^{\ast}$ on the left we get%
\begin{subequations}
\begin{align}
\dot{\theta}\left(  T\right)   &  =\frac{1}{iN_{0}\mu}\mathbf{w}_{0}^{\ast
}\mathcal{K\bar{B}}\boldsymbol{\xi}\left(  T\right)  ,\\
\dot{c}_{1}\left(  T\right)   &  =\lambda_{1}c_{1}\left(  T\right)
+\mathbf{w}_{1}^{\ast}\mathcal{K\bar{B}}\boldsymbol{\xi}\left(  T\right)
,\text{\ }%
\end{align}
where we have taken $c_{0}=0$ (this can be done because the arbitrary phase
$\theta$ can be conveniently redefined in order to collect the information on
this mode). We notice that although $\mathcal{K}$ and $\mathcal{\bar{B}}$
depend on phase $\theta$, $\mathcal{K\bar{B}}$ does not as can be checked from
Eqs. (\ref{F}) and (\ref{B}). Hence these equations are truly decoupled for
$\theta$ and $c_{1}$.

In the stationary limit, i.e. for large $T$, the solution of the above
equations reads%

\end{subequations}
\begin{subequations}
\begin{align}
\theta\left(  T\right)   &  =\theta\left(  0\right)  +\frac{\mathbf{w}%
_{0}^{\ast}\mathcal{K\bar{B}}}{iN_{0}\mu}\int_{0}^{T}dT^{\prime}\boldsymbol{\xi}%
\left(  T^{\prime}\right)  ,\label{fd}\\
c_{1}\left(  T\right)   &  =\mathbf{w}_{1}^{\ast}\mathcal{K\bar{B}}\int
_{0}^{T}dT^{\prime}\boldsymbol{\xi}\left(  T^{\prime}\right)  e^{-2\left(
T^{\prime}-T\right)  }. \label{c1}%
\end{align}

Finally we evaluate, for later purposes, the correlation spectrum of $c_{1}$.
From the formal solution (\ref{c1}), it is straightforward to prove that the
correlation function of this projection reads%
\end{subequations}
\begin{equation}
C_{1}\left(  \tau\right)  =\left\langle c_{1}\left(  T\right)  c_{1}\left(
T+\tau\right)  \right\rangle =\frac{\mathbf{w}_{1}^{\ast}\mathcal{K\bar{D}%
K}\mathbf{w}_{1}^{\ast}}{4}e^{-2\left\vert \tau\right\vert }.
\end{equation}
It is not difficult to obtain that $\mathbf{w}_{1}^{\ast}\mathcal{K\bar{D}%
K}\mathbf{w}_{1}^{\ast}=-4\kappa$, and then the spectrum of correlation
$C_{1}\left(  \tau\right)  $ turns out to be%
\begin{equation}
\tilde{C}_{1}\left(  \omega\right)  =\int_{-\infty}^{+\infty}d\tau
e^{-i\omega\tau}C_{1}\left(  \tau\right)  =-\frac{\kappa}{4+\omega^{2}}.
\label{CorrSpec}%
\end{equation}

\subsection{Dynamics of the bright mode's orientation}

We first analyze the dynamics of the output pattern orientation, governed by
$\theta$. Eq. (\ref{diffase}) shows that the phase $\theta$ diffuses with time
what means that the orientation of the classical mode in which emission
occurs, Eq. (\ref{Aclass}), exhibits a random walk. Then, although the mode
orientation is well defined at every instant, it can be understood that the
pattern orientation is undefined as after some time any value between $0$ and
$2\pi$ could be found. This is what we understand when we say that the
orientation of the output pattern is undetermined.%

It is important to see how much does $\theta$ diffuse. From Eq. (\ref{fd}) it
is straightforward to show that the variance of $\theta$ is given by%

\begin{subequations}
\begin{equation}
\left\langle \delta\theta\left(  T\right)  ^{2}\right\rangle =d_{\theta}T,
\label{diffase}%
\end{equation}
where we have used the notation $\delta A=A-\left\langle A\right\rangle $, and%
\end{subequations}
\begin{align}
d_{\theta}  &  =-\frac{\mathbf{w}_{0}^{\ast}\mathcal{K\bar{D}K}\mathbf{w}%
_{0}^{\ast}}{N_{0}^{2}\mu^{2}}\nonumber\\
&  =\kappa\frac{\mu^{2}\sin2\phi_{0}+p\sin\left[  2\left(  \phi_{0}%
+\psi\right)  \right]  }{4\mu^{2}\cos^{2}\phi_{0}},
\end{align}
with $\psi$, $\mu$, and $\phi_{0}$ given by Eqs. (\ref{psi}), (\ref{mu}) and
(\ref{fi0}).

In Fig. 3 we represent $D\equiv d_{\theta}/\kappa$ as a function of the pump
strength $p$ for several values of $\Delta$. Notice that for $\mu\rightarrow0$
(i.e., at the supercritical bifurcation that occurs in the upper branch of the
domain of existence of (\ref{solON}), see Fig. 2), $D\rightarrow\infty$. This
is an intuitive result because when the output mode mean photon number is
close to zero, the pattern orientation can be abruptly changed with the
addition of a single couple of photons. As the system is brought apart from
this bifurcation the mean number of photons rapidly increases and,
consequently, it is more difficult for the fluctuation to change the
orientation of the pattern, what is obviously consistent with the rapid
decrease of $D$. As for the quantitative value notice that, except very close
to the supercritical bifurcations, $D$ is a quantity of order one what means
that the diffusion constant $d_{\theta}$ is basically of the same order of
magnitude as $\kappa=g/\gamma_{s}$, which we have assumed to be a very small
number. Consequently, although the TEM\ output mode is randomly rotating in
the transverse plane, the rotation is very slow except very close to the
supercritical bifurcation points.

\begin{figure}[t]

\includegraphics[
height=2.3229in,
width=3.237in
]%
{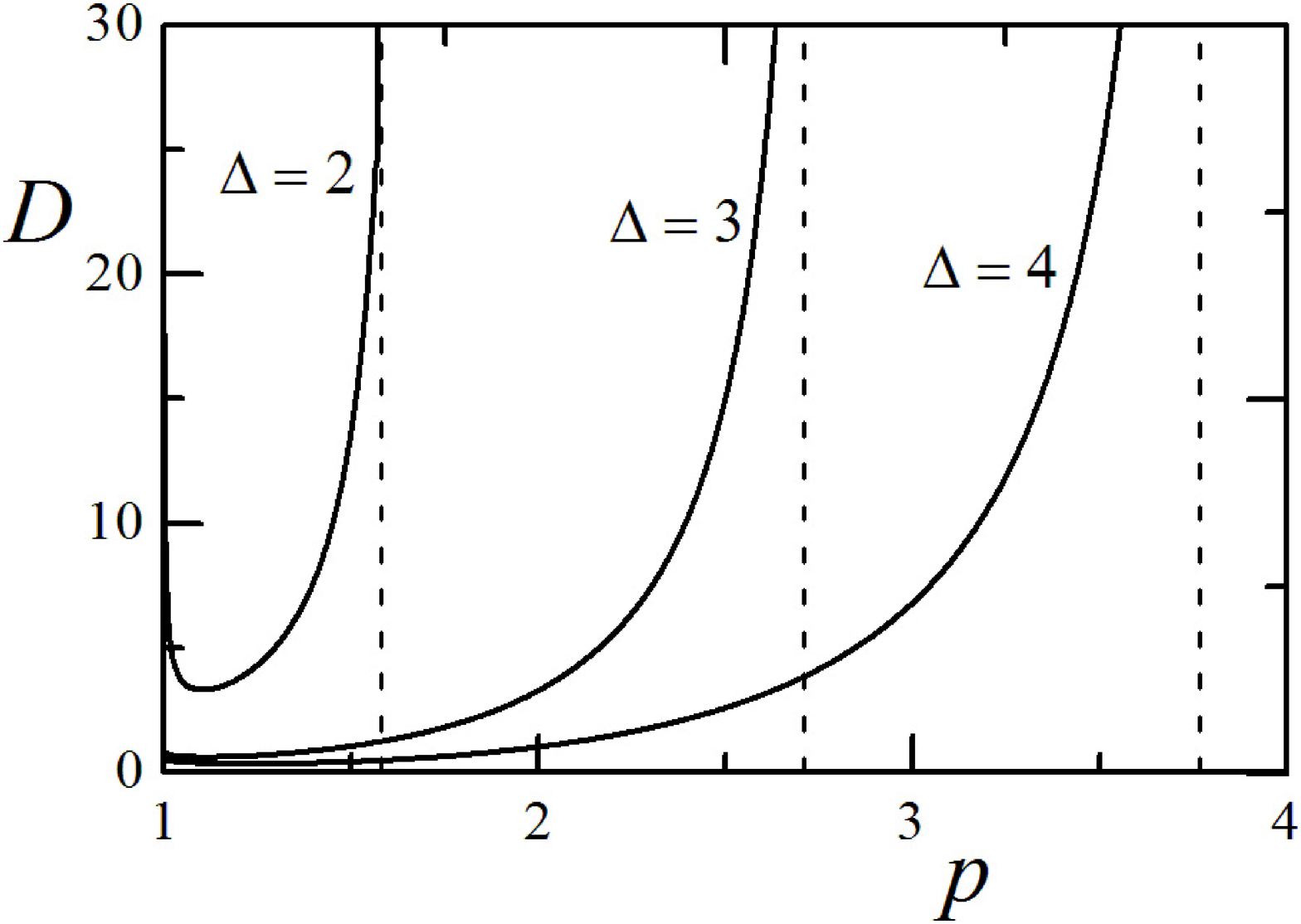}%
\\
{\protect\small Figure 3.- Dependence of the normalized diffusion constant
$D=d_\theta/\kappa$ as a function of the pump strength $p$ for the
values of the detuning indicated in the figure. The vertical dashed lines
indicate the values of $p$ where the nontrivial solution ceases to exist.
}%

\end{figure}

\subsection{Non-critical squeezing properties of the dark mode}

Now we focus on the main result of the present article: We show below that the
dark mode has complete noise reduction on its phase quadrature irrespective of
the system parameters. To this aim we evaluate its quadrature fluctuations as
measured in an homodyne detection experiment. As it is well known (see, e.g.,
\cite{Gea-Banacloche}), quadrature fluctuations outside the cavity are given
by the noise spectrum, which for a general problem quadrature $\hat{X}%
_{m}^{\varphi}$ ($m$ refers to any of the transverse modes of our system) is
given by%
\begin{equation}
V^{out}\left(  \omega;X_{m}^{\varphi}\right)  =1+S_{m}^{\varphi}\left(
\omega\right)  , \label{Vout}%
\end{equation}
with $S_{m}^{\varphi}\left(  \omega\right)  $ the squeezing spectrum that,
after taking into account our rescaling (\ref{reescal}), can be written as%
\begin{equation}
S_{m}^{\varphi}\left(  \omega\right)  =\frac{2}{g}\int_{-\infty}^{+\infty
}d\tau e^{-i\omega\tau}\left\langle \delta X_{m}^{\varphi}\left(  T\right)
\delta X_{m}^{\varphi}\left(  T+\tau\right)  \right\rangle . \label{Sout}%
\end{equation}
Defined in this way, $V^{out}\left(  \bar{\omega}\right)  =0$ means complete
absence of fluctuations at $\omega=\bar{\omega}$, while $V^{out}\left(
\bar{\omega}\right)  =1$ means that fluctuations at $\omega=\bar{\omega}$ are
those corresponding to the vacuum state.

In the homodyning, the spatial profile of the local oscillator field (LOF)
selects the transverse mode to be measured, while its phase selects a
particular mode's quadrature. In what follows we suppose that the LOF is
perfectly matched to the \textit{dark} mode's profile $H_{s}^{\theta}\left(
\mathbf{r}\right)  $. As we saw in (\ref{XtoC}), the independent quadratures
of this mode are $X_{s,\theta}^{\phi_{0}}$ and $X_{s,\theta}^{\pi/2}$, from
which we can build a general quadrature as%
\begin{equation}
X_{s,\theta}^{\varphi}=\frac{1}{\cos\phi_{0}}\left[  X_{s,\theta}^{\phi_{0}%
}\cos\varphi+Y_{s,\theta}\sin\left(  \varphi-\phi_{0}\right)  \right]  .
\end{equation}
This expression is readily obtained from the more common expressions
$X_{s,\theta}^{\varphi}=X_{s,\theta}^{0}\cos\varphi+X_{s,\theta}^{\pi/2}%
\sin\varphi$ and $X_{s,\theta}^{\phi_{0}}=X_{s,\theta}^{0}\cos\phi
_{0}+X_{s,\theta}^{\pi/2}\sin\phi_{0}$.

Hence, by using the relation between the independent quadratures and the
projections $c_{j}$ (\ref{XtoC}), and remembering Eq. (\ref{CorrSpec}) and
that $c_{0}=0$, it is trivial to find the noise spectrum of this general
quadrature, which reads
\begin{align}
V^{out}\left(  \omega;X_{s,\theta}^{\varphi}\right)   &  =\frac{\cos
^{2}\varphi+\sin^{2}\left(  \varphi-\phi_{0}\right)  }{\cos^{2}\phi_{0}}\\
&  -\frac{\sin^{2}\left(  \varphi-\phi_{0}\right)  }{\cos^{2}\phi_{0}}\frac
{1}{1+\left(  \omega/2\right)  ^{2}}.\nonumber
\end{align}
This expression shows that the quantum properties of the dark mode of the
current system are exactly the same as that found in \cite{PRL, Largo} for the
case of a DOPO cavity: at $\omega=0$, $V^{out}\left(  \omega=0;X_{s,\theta
}^{\varphi}\right)  =\cos^{2}\varphi/\cos^{2}\phi_{0}$ and thus it has
complete absence of fluctuations on its phase quadrature ($\varphi=\pi/2$),
while another of its quadratures ($\varphi=\phi_{0}$ in our case) carries only
with vacuum fluctuations. Any other quadrature having $\varphi$ between
$\pi/2$ and $\phi_{0}$ is squeezed below the vacuum level, though the
squeezing level is smaller as $\varphi$ approaches to $\phi_{0}$. These
results are independent of the system parameters, what we expected as the
noise reduction relies on the rotational symmetry breaking only.

It could seem that this result violates the uncertainty principle, as the
product of the noise spectra corresponding to two orthogonal quadratures is
below unity. However, in \cite{Largo}, we have proven that this is actually
not the case, as the canonical pair of the squeezed quadratures is not another
quadrature, but the orientation of the dark mode $\theta$, which is indeed
undetermined in the long time term.

\section{Conclusions}

We have proposed a model for a Kerr cavity in which a spontaneous rotational
symmetry breaking occurs when the system is beyond the emission threshold: The
nonlinear cavity has a perfect rotational symmetry and is pumped by Gaussian
beams, but the emitted signal field has the shape of a TEM$_{10}$ mode that
breaks the rotational symmetry. We have demonstrated in a special simple limit
(in which the pumping fields are taken as constants) that the rotational
symmetry breaking implies (i)\ the diffusion of the output mode orientation,
and (ii) the perfect squeezing of the phase quadrature of the TEM$_{10}$ mode
that is rotated $\pi/2$ with respect to the signal TEM$_{10}$ mode. These
results are in perfect agreement with our previous proposal of the symmetry
breaking mediated squeezing in a DOPO\ model \cite{PRL, Largo}. The interest
of the results here presented are twofold. On one hand, we are proposing a
system different to that of \cite{PRL, Largo} for the possible observation of
the phenomenon, thus showing that the results in \cite{PRL, Largo} are quite
general. On the other hand, rotational symmetry could be broken in a
DOPO\ cavity if angular phase matching is necessary, as in this case the
nonlinear crystal axis is rotated a certain angle with respect to the cavity
axis, a problem that does not exist in the case of a $\chi^{\left(  3\right)
}$ nonlinear medium as phase matching is easier to obtain.

As for the particular model we have proposed, in its formulation we have
assumed a confocal cavity, as in this cavity type the required resonances are
verified. Notice however that the ingredients that are essential for the
phenomenon of rotational symmetry breaking mediated squeezing generation are:
(i) rotational invariance, and (ii)\ that the signal field photons have
non--null OAM. Then, the dynamics of the pumping modes is irrelevant except
for the quantitative details. In this sense, the use of a confocal cavity is
not essential and any other $\chi^{\left(  3\right)  }$ cavity in which the
signal modes are the right ones could exhibit the described phenomenon. Then
this phenomenon could possibly be observed in other $\chi^{\left(  3\right)
}$ cavities such as, e.g., fiber ring resonators.

Another comment we would like to add concerns the non--pump depletion
approximation. How would the dynamics of the pumping modes affect the results
we have derived? The inclusion of the pumping fields equations in the study
would obviously introduce more eigenvalues (the ones corresponding to the
stability of these modes) and would modify some of the eigenvalues governing
the dynamics of the signal modes, but would not modify the existence of a
Goldstone mode once the signal field is switched on (as it appears due to the
symmetry breaking). Hence as far as the dynamics of the pumping modes does not
destroy completely the stability of the cw signal field emission, wherever the
signal modes are stable the phenomenon will be present. It could well happen
that the general model exhibits Hopf bifurcations that would reduce the domain
of stable existence of the cw signal modes, but there will be a finite domain
of stability for these modes and within it there will be the perfect squeezing
properties we have described in our work.

We would finally stress two important features. In \cite{PRL} we demonstrated
for a DOPO\ model that small imperfections in the rotational symmetry do not
lead but to a small degradation of the squeezing level. On the other hand, in
\cite{Largo} we have numerically demonstrated for the same DOPO\ model that
the rotational symmetry breaking mediated squeezing is perfect beyond the
linear approximation. These conclusions should also hold for the system here presented.

\appendix
The components of the drift vector in Eq. (\ref{FP}) read
\begin{subequations}
\label{Alan}%
\begin{align}
A_{\alpha_{1}}  &  =\mathcal{E}_{p}-\left(  \gamma_{1}+i\delta\right)
\alpha_{1}+4ig\alpha_{2}^{+}\alpha_{2}\alpha_{1}+\label{A1}\\
&  2ig\left(  \alpha_{1}^{+}\alpha_{1}^{2}+\alpha_{+}^{+}\alpha_{+}\alpha
_{1}+\alpha_{-}^{+}\alpha_{-}\alpha_{1}+\alpha_{2}^{+}\alpha_{+}\alpha
_{-}\right)  ,\nonumber\\
A_{\alpha_{2}}  &  =\mathcal{E}_{p}-\left(  \gamma_{2}+i\delta\right)
\alpha_{2}+4ig\alpha_{1}^{+}\alpha_{1}\alpha_{2}+\label{A2}\\
&  2ig\left(  \alpha_{2}^{+}\alpha_{2}^{2}+\alpha_{+}^{+}\alpha_{+}\alpha
_{2}+\alpha_{-}^{+}\alpha_{-}\alpha_{2}+\alpha_{1}^{+}\alpha_{+}\alpha
_{-}\right)  ,\nonumber\\
A_{\alpha_{+}}  &  =-\left(  \gamma_{s}+i\delta\right)  \alpha_{+}%
+ig\alpha_{+}^{+}\alpha_{+}^{2}+\\
&  2ig\left(  \alpha_{-}^{+}\alpha_{-}\alpha_{+}+\alpha_{1}^{+}\alpha
_{1}\alpha_{+}+\alpha_{2}^{+}\alpha_{2}\alpha_{+}+\alpha_{-}^{+}\alpha
_{1}\alpha_{2}\right)  ,\nonumber\\
A_{\alpha_{-}}  &  =-\left(  \gamma_{s}+i\delta\right)  \alpha_{-}%
+ig\alpha_{-}^{+}\alpha_{-}^{2}+\\
&  2ig\left(  \alpha_{+}^{+}\alpha_{+}\alpha_{-}+\alpha_{1}^{+}\alpha
_{1}\alpha_{-}+\alpha_{2}^{+}\alpha_{2}\alpha_{-}+\alpha_{+}^{+}\alpha
_{1}\alpha_{2}\right)  ,\nonumber
\end{align}
and the rest of components $A_{\alpha_{k}^{+}}$ are as $A_{\alpha_{k}}$ after
complex--conjugating and swapping $\alpha_{k}\ $and $\alpha_{k}^{+}$. As for
the diffusion matrix in Eq. (\ref{FP}), it reads%
\end{subequations}
\begin{equation}
\mathcal{D}=2ig%
\begin{pmatrix}
\mathcal{D}^{\left(  -\right)  } & 0\\
0 & -\mathcal{D}^{\left(  +\right)  }%
\end{pmatrix}
,
\end{equation}
being $\mathcal{D}^{\left(  -\right)  }$ a 4$\times$4 matrix with elements
\begin{subequations}
\label{Dlan}%
\begin{align}
\mathcal{D}_{11}^{\left(  -\right)  }  &  =\alpha_{1}^{2},\mathcal{D}%
_{22}^{\left(  -\right)  }=\alpha_{2}^{2},\mathcal{D}_{33}^{\left(  -\right)
}=\frac{\alpha_{+}^{2}}{2},\mathcal{D}_{44}^{\left(  -\right)  }=\frac
{\alpha_{-}^{2}}{2},\\
\mathcal{D}_{12}^{\left(  -\right)  }  &  =\mathcal{D}_{21}^{\left(  -\right)
}=2\alpha_{1}\alpha_{2}+\alpha_{+}\alpha_{-},\\
\mathcal{D}_{13}^{\left(  -\right)  }  &  =\mathcal{D}_{31}^{\left(  -\right)
}=\alpha_{1}\alpha_{+},\mathcal{D}_{14}^{\left(  -\right)  }=\mathcal{D}%
_{41}^{\left(  -\right)  }=\alpha_{1}\alpha_{-},\\
\mathcal{D}_{23}^{\left(  -\right)  }  &  =\mathcal{D}_{32}^{\left(  -\right)
}=\alpha_{2}\alpha_{+},\mathcal{D}_{24}^{\left(  -\right)  }=\mathcal{D}%
_{42}^{\left(  -\right)  }=\alpha_{2}\alpha_{-},\\
\mathcal{D}_{34}^{\left(  -\right)  }  &  =\mathcal{D}_{43}^{\left(  -\right)
}=\alpha_{1}\alpha_{2}+\alpha_{+}\alpha_{-},
\end{align}
\end{subequations}
and $\mathcal{D}^{\left(  +\right)  }$ as $\mathcal{D}^{\left(  -\right)  }$
after swapping $\alpha_{k}$ and $\alpha_{k}^{+}$.

%

\end{document}